# Flexible Physical Unclonable Functions based on non-deterministically distributed Dye-Doped Fibers and Droplets

Mauro Daniel Luigi Bruno[1,2,*], Giuseppe Emanuele Lio[3,4,5,*,⊛], Antonio Ferraro[2,1,*,⊛], Sara Nocentini[6,4], Giuseppe Papuzzo[7], Agostino Forestiero[7], Giovanni Desiderio[2], Maria Penelope De Santo[1,2], Diederik Sybolt Wiersma[3,4,6], Roberto Caputo[1,2,8], Giovanni Golemme[9], Francesco Riboli[5,4], and Riccardo Cristoforo Barberi[1,2]

1  *University of Calabria, Physics Department, 87036 Arcavacata di Rende (CS), Italy*
2  *Consiglio Nazionale delle Ricerche - Istituto di Nanotecnologia CNR-Nanotec, Rende (CS), 87036 Italy*
3  *Physics Dept., University of Florence, Via Sansone, 1, 50019, Sesto Fiorentino, Florence, Italy*
4  *European Laboratory for Non-Linear Spectroscopy (LENS), University of Florence, Via Nello Carrara 1, 50019, Sesto Fiorentino, Florence, Italy*
5  *Consiglio Nazionale delle Ricerche - National Institute of Optics, CNR-INO, 50019, Sesto Fiorentino (FI), Italy*
6  *Istituto Nazionale di Ricerca Metrologica (INRiM), Strada delle Cacce 91, Turin, 10135 Italy*
7  *Institute for High Performance and Networking, CNR-ICAR, via P. Bucci 8-9c, 87036 Rende, Cosenza, Italy*
8  *Institute of Fundamental and Frontier Sciences, University of Electronic Science and Technology of China, Chengdu 610054, China*
9  *University of Calabria, Environmental Engineering Department, 87036 Arcavacata di Rende (CS), Italy*

⊛ *Authors contribute equally to this work.*

The development of new anti-counterfeiting solutions is a constant challenge and involves several research fields. Much interest is devoted to systems that are impossible to clone, based on the Physical Unclonable Function (PUF) paradigm. In this work, new strategies based on electrospinning and electrospraying of dye-doped polymeric materials are presented for the manufacturing of flexible free-standing films that embed different PUF keys. Films can be used to fabricate anticounterfeiting labels having three encryption levels: i) a map of fluorescent polymer droplets, with non deterministic positions on a dense yarn of polymer nanofibers; ii) a characteristic fluorescence spectrum for each label; iii) a challenge-response pairs (CRPs) identification protocol based on the strong nature of the physical unclonable function. The intrinsic uniqueness introduced by the deposition techniques encodes enough complexity into the optical anti-counterfeiting tag to generate thousands of cryptographic keys. The simple and cheap fabrication process as well as the multilevel authentication makes such colored polymeric unclonable tags a practical solution in the secure protection of merchandise in our daily life.

## I. INTRODUCTION

Today, there is a growing interest in developing anti-counterfeiting systems to protect goods and the health of consumers. Counterfeit banknotes, certificates, medicine, and electronic components are constantly found, leading to serious economic losses and even endangering in some cases human health [1–7]. This issue affects also the textile and food industry where often false statements of origin are seen on products. Barcode, QR-code, fluorescent ink, and DNA marking is being used as anti-counterfeiting device [8], although their production process is deterministic and can, therefore, be cloned easily [9]. Recent technologies based on plasmonic security tags, bio-mimetic micro-fingerprints and magnetic responses achieved high levels of protection against cloning attacks [10, 11]. The paradox is, however, that many of the stronger anti-counterfeiting systems can be more expensive than the protected products. Moreover, specific and expensive tools are usually required to verify their operation. Lower cost solutions, which offer improved security, can be attained by exploiting the luminescence of organic dyes, carbon dots, and semiconductor nanoparticles. However, their long-term toxicities or broad emission bands hinder their further implementation [12–18].

Researchers aim to develop anti-counterfeiting labels that are easy to be validated, but impossible to be cloned also for the manufacturers. One of the best solutions is represented by embedding a Physical Unclonable Function (PUF) key in the label that, when interrogated with a specific input (challenge), returns a unique, unpredictable output (response) [19].

During the last decade, many efforts have been undertaken to produce strong and attack resistant PUFs using a broad combination of materials and physical phenomena. A wide portfolio is represented by organic crystals, photonic crystals, fluorescent materials, perovskites, metamaterials, soft materials, and polymers, that offer peculiar optical properties and random morphologies led by self-assembling [10, 11, 20–32]. Recently, electrospinning has emerged as a technique for the realization of anti-counterfeiting tags. In 2018, Gangwar and co. [33] fabricated a dual mode flexible and luminescent white security paper and nano-taggants based on polymeric nanofibers and nanophosphors. This tag, white under the daylight, can show a strong red or green color depending on the excitation wavelength source used. The same approach is used in [34] where the film tag does not show any feature under white light but hidden information can be visualized under UV light. Wu and co. [35] proposed the fabrication of rewritable paper as anti-counterfeiting film based on dual-stimuli responsive color-changing nanofibrous membranes prepared by electrospinning. The paper discoloration occurs through thermal or solvent stimuli. A similar technique, namely electrospraying, is used for the fabrication of unclonable tag by generating micrometer scale features with random position [36].

Herein, we exploit electrospinning and electrospraying techniques for the realization of new anti-counterfeiting tags based on Physical Unclonable Functions. Both procedures rely on the injection, through a glass syringe, of a polymer solution while applying a strong electric field between the syringe conductive needle (cathode) and a metallic collector (anode) placed few centimeters one from the other. When the polymer solution is injected, at the needle apex a droplet is formed, whose surface charge increases as a function of the applied electric field. This causes the droplet deformation that changes its shape into a conical one, the so-called Taylor cone. Once the applied potential overcomes a threshold voltage a filament is expelled from the cone towards the collector. The solvent quickly evaporates and dried fibers deposit on the collector surface. The difference between electrospinning and electrospraying technologies lies in the viscosity of the polymer solution, and so on its concentration and molecular weight. For high polymer concentration, a yarn is produced [37, 38]. For low concentrations, droplets, instead of filaments, are formed [39–41]. The injection flow rate and the applied voltage also play an important role: low voltages promote droplets instead of fibers formation. In this work anti-counterfeiting tags are realized by electrospraying fluorescent dye-doped polymer microdroplets on a dense layer of electrospun fluorescent dye-doped polymer nanofibers. The process is completely non-deterministic, the position of fibers as well as of droplets is impossible to control and reproduce representing the first security level. It is worth noting that for the experiments only water is used as solvent for the polymer thus enabling an eco-friendly production. The fluorescence emission of the single label represents the second security level. Each label, containing a different amount of the used fluorophores, has its own fluorescence spectrum. The third strong security level arises from the unique morphology of the nanofiber layer and the microdroplets which produce a unique speckle pattern when illuminated with coherent light such as laser light. The proposed strategy satisfies the basic requirements to produce an anti-counterfeiting label: unpredictable, unclonable, use of low cost materials, easy to produce, and inexpensive authentication systems. Moreover, thanks to their high flexibility, the realized labels can easily applied on every kind of good and merchandise.

## II. ENCRYPTION LEVELS

The proposed security tags are made of nanofibers and microdroplets doped with different fluorescent dyes as shown in Figure 1a. A UV fluorescent dye, S420, is used for the polymer nanofibers, the emission of which overlaps the absorption of Rhodamine B, the dye used for the microdroplets. The authentication process initially consists in the full characterization of the tag for the three security levels. Once the labels have been produced, they are illuminated with a UV lamp and photographed with different rotation angles, in order to guarantee the authentication in all the conditions. A smartphone is used for this procedure. Then, the unique map of scatterers points is obtained using a developed recognition algorithm and stored into a secure server (central authority) that will be consulted at the time of validation, see Figure 1a.

Then, the unique fluorescence emitted by each tag arising from the non deterministic position and composition of microdroplets, is recorded, as shown in Figure 1b. The authentication procedure is similar to the one described for the first level except for the different nature of data stored in the secure server which now are fluorescence spectra. They can be acquired using a portable spectrophotometer for smartphones, or a laboratory one. Portable spectrophotometers can be purchased for a few hundred dollars, an affordable expense to the consumer and for small and medium-sized enterprises. The user, the seller or the buyer, that would like to authenticate the merchandise, interrogates the label with a subset of the available challenges and, if the collected responses match the recorded ones, the access is granted. Otherwise, it is denied. It is worth noting that both security levels can be authenticated with the use of a smartphone using ad-hoc developed apps: this is an important result

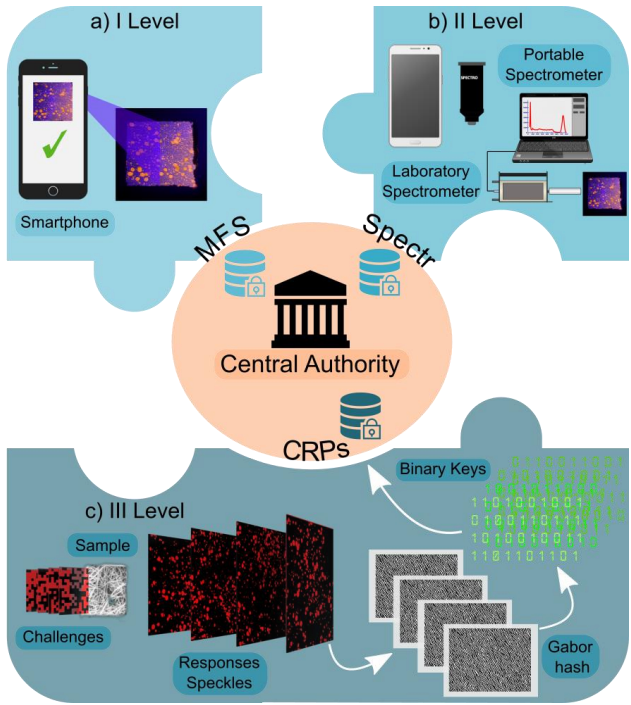

FIG. 1. a) The sketch shows the validation workflow of merchandise labeled with the proposed multi-level PUF/authentication scheme key based anti-counterfeiting label. The consumer can accesses to the first two identification levels by using a smartphone (a) for analyzing the unique map of fluorescence scatterers in combination with an UV lamp; (b) and for analyzing the fluorescence emission of the label in combination with a portable spectrophotometer. Laboratory spectrophotometer can be also used; (c) Finally, the third level consists in a forensic analysis made by exploiting the speckle pattern produced by the label when investigated with the challenge-response pairs scheme. For all three levels, the authentication of each label is performed by comparing the onsite response with the one stored by the manufacturer into a central authority server.

## III. RESULTS

To fabricate the labels, a dye doped polyvinyl alcohol(PVA)/water solution is prepared with the following percentages by weight: $99.7\%(92\% H_2O + 8\% PVA) + 0.3\%(S420)$. Then, the solution is electrospun using a high voltage supply and a syringe pump. The liquid solution is injected through a syringe and once reached the tip of the needle, a fiber starts to be formed following the trajectory of the Taylor cone: the results is a white yarn of fibers distributed on the collector plane. The fluorescent dye S420 cannot be distinguished by eye because its absorption band is 350nm with a tail up to 400nm while the emission peak is at 420nm. Hence it is possible to distinguish a intense blue color if illuminated with a UV lamp (Figure 2a). Another solution is prepared replacing S420 with Rhodamine B that possesses an absorption band overlapped with the emission band of the S420 dye. The solution is then electrosprayed above the fibers layer, using the same experimental setup, setting a lower voltage and higher flow rate with respect to electrospinning process. The droplets can be partially distinguished by eye due to the use of a fluorescent dye which absorbs in the visible range. However, by illuminating the two dyes PUF typology with the same UV lamp, droplets are distinguished showing overall more brilliant colors, see Figure 2. The use of dye S420 as background enhance the brightness of Rhodamine B (S1). The mechanical stability of the labels depends on its thickness and it is not very high, but it can be improved sandwiching the label between two adhesive transparent films destructible upon removal without affecting the anti-counterfeiting capacity. The final results is a polymer substrate of $15cm \times 15cm$. From a single batch of fabrication, it is possible to obtain more anti-counterfeting labels of small dimensions, as shown in Figure 2c, possessing unique and unpredictable features. It is worth noting that the cloning of such tag is impossible due to its not deterministic deposition of fibers and droplets. Once the labels are produced, pictures of

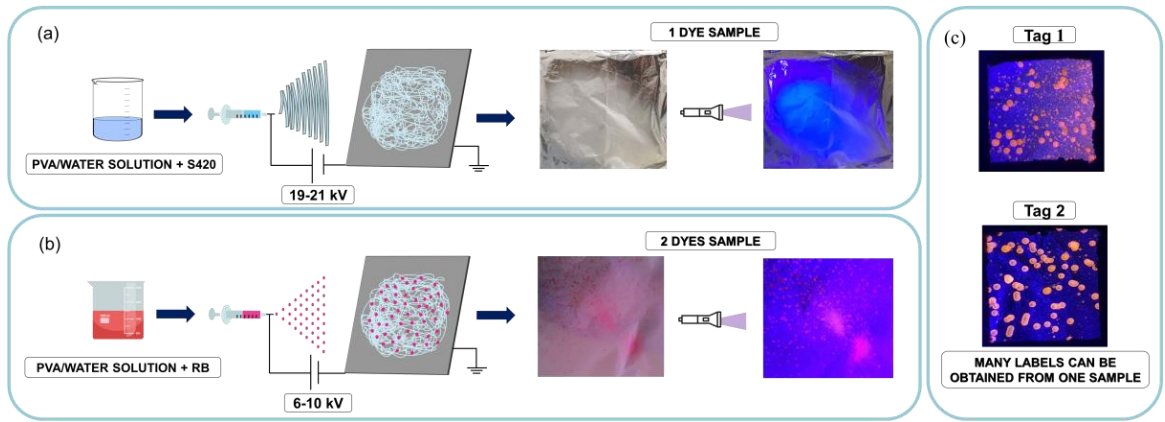

FIG. 2. (a) Electrospun polyvinyl alcohol (PVA)/water solution doped with fluorescent dye S420. Illuminating the fibers yarn with a UV lamp (λ=370nm), a strong blue color can be observed. (b) The electrospray of the PVA/water solution doped with Rhodamine B is performed on the fibers yarn shown in (a). The fluorescence emission of the two dyes can be observed illuminating the labels with UV light (λ=370nm). The size of the aluminum foil collector is $15cm \times 15cm$. (c) From one single fabrication step, it is possible to obtain many labels ($2cm \times 2cm$).

each tag are taken and all stored in a server. Therefore, photographs of each label at different rotation angle is acquired showing the position of the microdroplets. Then, the acquired image is processed and compared with those stored in a server through an appositely developed app for mobile devices, illuminating the tag with a UV source of light. The application identifies the position of all the microspheres and creates a map of points relating each microsphere to the others by their respective distances. Identical recognized microspheres are linked with a line and for the authentication a large amount of lines is necessary. Furthermore, to facilitate the reading of the tag during identification and authentication, the algorithm has been developed in such a way that, even rotating the tag by several degrees it is still possible to make a comparison between tags. Figure 3a shows the correlations between the original stored picture of a tag with itself (3a (i)) and with the same image rotated of 90° (3a (ii)), 180° (3a (iii)). As shown in the graph in Figure 3a, the algorithm recognizes the 100% of the features comparing the same images of the tag, and when the tag is rotated, the recognition percentage lowers, but does not go below the 20%. In the situation in which the label is not stored, i.e. a fake label, the algorithm recognizes only few points that can be considered "false positive". As we can see in Figure 3b (i), few lines connect only three points between two uncorrelated images. The recognition percentage decreases rotating the tag of 90° and 180° (3b (ii)). The graph in Figure 3(b) shows the recognition percentage achieved with the rotation angle. The number of the recognized features is low or zero so that it can be assumed that the images are not the same and we are in presence of a counterfeited tag. To strengthen the security of the proposed labels, it is possible to exploit their fluorescence emission by simply using portable spectrophotometers such as those present in a laboratory or one of the modern smartphone spectrophotometers with open source applications. Fluorescent spectrum varies in each label, depending on the dyes employed and on the the amount of fibers and droplets deposited. The labels shown in the Figure 2c are an example: Tag1 has a higher quantity of droplets containing Rhodamine B, while Tag2 contains less droplets and shows a more intense blue color, due to dye S420. Using a UV light source emitting at a wavelength of 310nm, Tag1 shows a strong pink colour due to Rhodamine B, in fact, the peak of this fluorescent dye is more intense than S420 while Tag2 presents the reverse situation showing an intense blue colour, see Figure 3c,d respectively. The fluorescence of Rhodamine B cannot be stimulated directly with the light source, so its emission must be stimulated, through energy transfer, by the fluorescent emission of S420. So, a single source of light can be used to stimulate the fluorescence of both dyes. It can be assumed that the relative fluorescence intensity of the peaks, one with respect to the other, of the different dyes, is unique for each tag. The challenge response pairs (CRPs) analysis is proposed as final step in the authentication process. It consists in investigating each tag with a coherent light ($\lambda$ =633nm that is shaped in 2000 different wavefronts using a DMD (defined as challenges), and then it impinges on the tag producing speckle patterns in far-field collected with a CCD camera (responses). Therefore, for each challenge-response pair an authority database is generated enrolling the collected CRPs that will be used in future for speckle comparison into the authentication process [47, 48], more details are available in the "Materials and Methods" section.) The first analysis is carried out on electrospun fibers, whose peculiar structure is investigated with scanning electron microscope

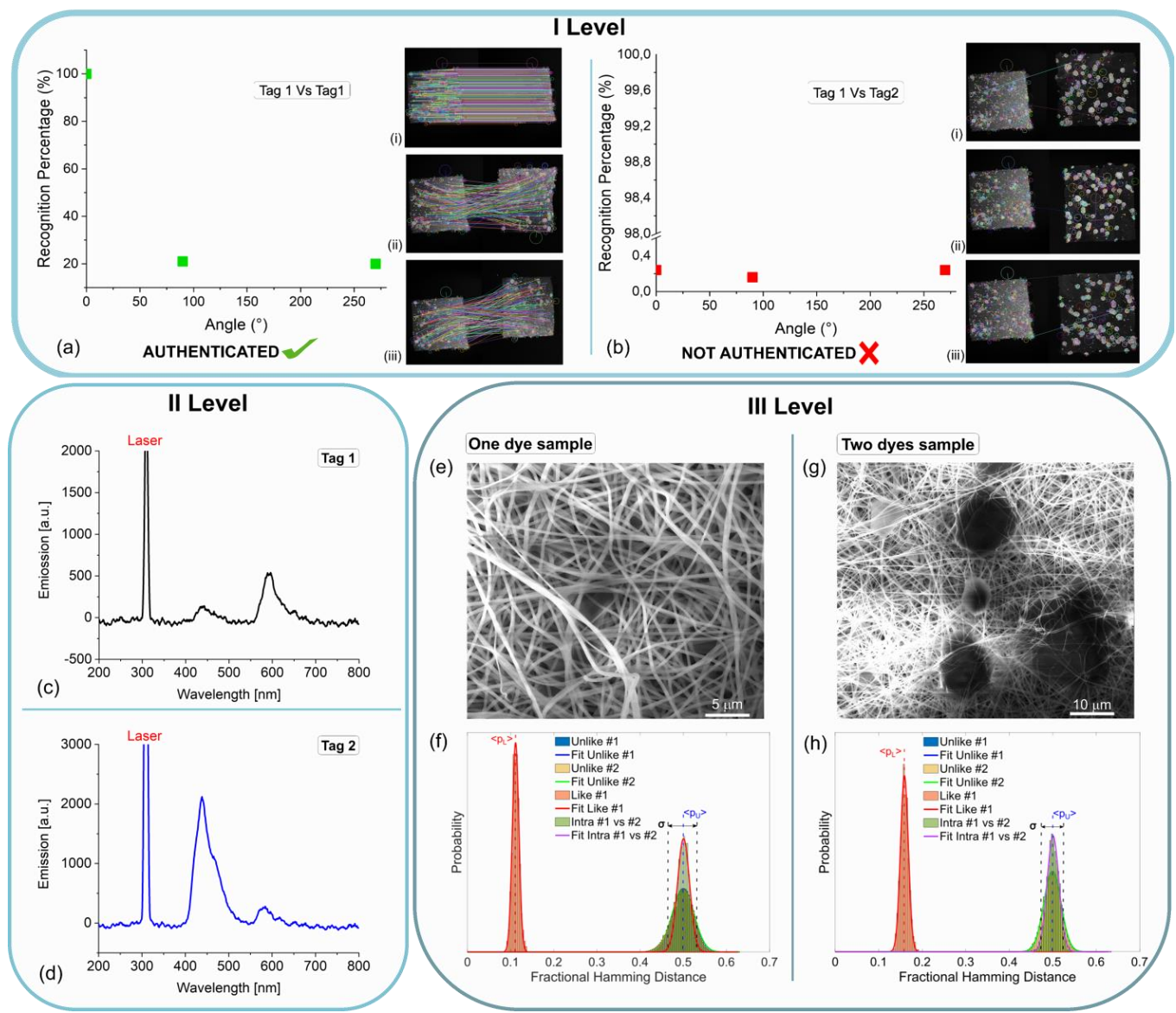

FIG. 3. Encryption levels. I level: analysis of the scatterers position. (a) A label (Tag1) is compared to an image of itself not rotated (i), rotated 90° (ii) and 180° (iii). The graph shows the recognition percentage versus the rotation angle that is 100% when the rotation angle is zero. Even rotating the label, the recognition percentage does not go below 20%. (b) A label (Tag1) is compared with the image of a different non-rotated label (Tag2) (i), rotated 90° (ii) and 180° (iii). The graph shows the recognition percentage that goes below 1% when the rotation angle is zero. Rotating the tag, the recognition percentage continues to decrease. II level: fluorescence analysis by illuminating the tags with a light source at 310nm. (c) Tag 1 shows a more intense Rhodamine B fluorescence peak than S420 ; (d) Tag 2 shows a more intense S420 fluorescence peak than Rhodamine B. SEM images of (e) polymer fibers obtained through the electrospinning of a PVA/water solution doped with S420. f) The speckle analysis (III level) done on two different pieces of the same PUF presents unlike FHD distributions centered at 0.5 and the like one at 0.12, blue, yellow and orange histograms respectively. The intra-device FHD, comparison of the two pieces of the same PUF, is centered at 0.5, light green histogram. (g) Polymer droplets of a PVA/water solution doped with Rhodamine B obtained through electrospray on a dense yarn of electrospinned polymer fibers containing S420. h) The FHD distributions on two pieces of this second PUF have "unlike" and "like" distributions centered at 0.5 and 0.16, while the intra-device one goes at 0.5.

(SEM), shown in Figure 3e. The CRPs analysis conducted on two different tags, from the same challenges set reveals an "unlike" distributions centered at $\mu$=0.5. Then, if two labels are compared the "intra" distributions are also centered at $\mu$=0.5 meaning the high unrelated level between each labels, see Figure 3f. The like distributions, instead, are centered at $\mu_l \sim$ 0.12 meaning that the system is stable. On the other hand, the inclusion of dispersed dye droplets (Rhodamine B) (by electrospray) into the fiber spun increases the morphology uniqueness of the proposed labels, see SEM image in Figure 3g. This new microscopic modification leads to more complex and unbreakable PUF. In fact, the results reported in Figure 3h confirm again the high difference among two labels produced with this technique. In this case the like distributions are centered at $\mu_l \sim 0.17$ meaning that the system is still stable, and it grants a FHD threshold value below of 0.2. Pearson correlation coefficient (PCC) has been adopted as an alternative metric to evaluate the differences or similarities in the analyzed tags [21], see Figure S4(a,b) in SI.

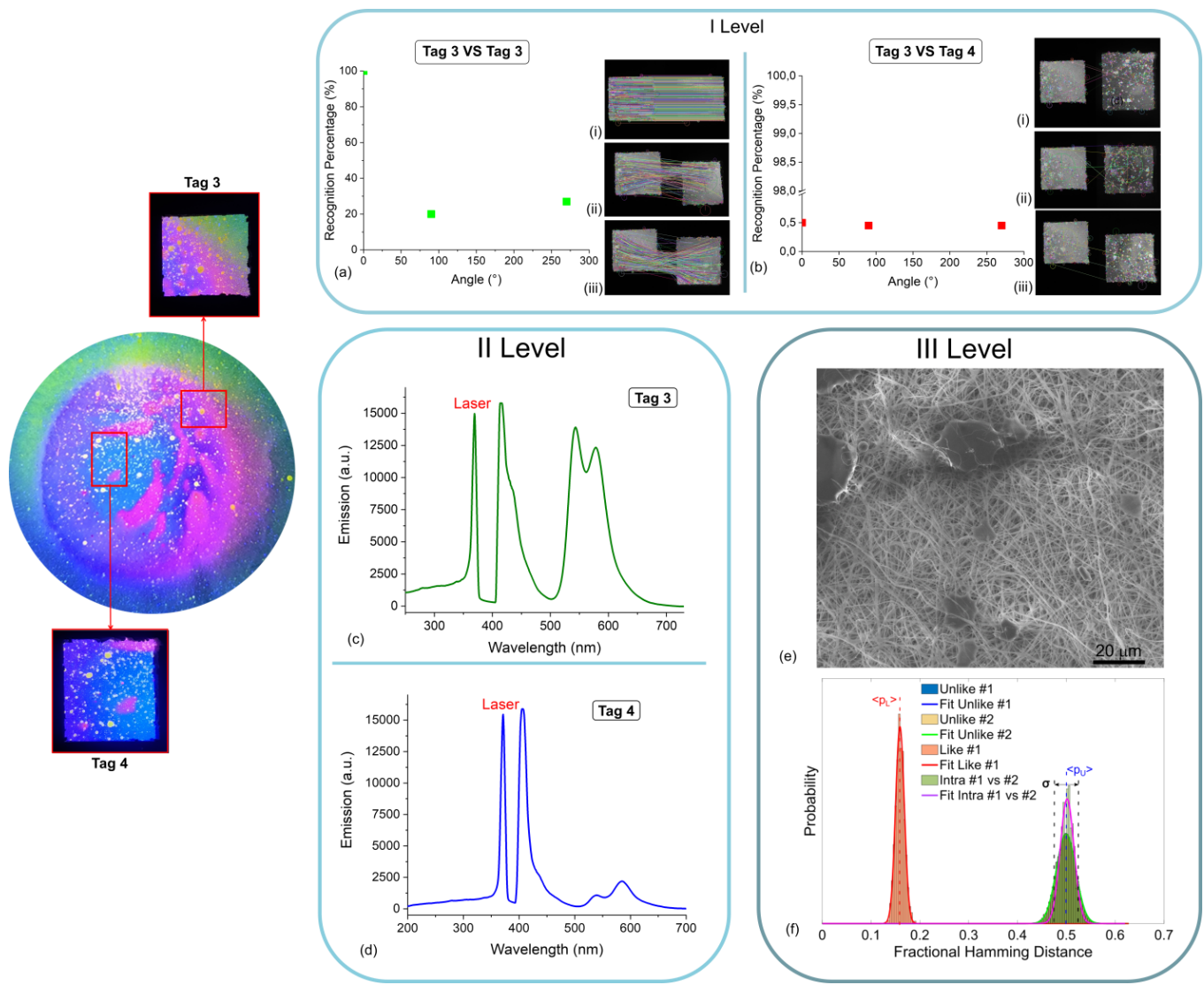

FIG. 4. Encryption levels. I level: analysis of the position of the scatterers are reported. (a) A label (Tag 3) is compared to an image of itself not rotated (i), rotated 90° (ii) and 180° (iii). The graph shows the recognition percentage achieved with the rotation angle that is 100% when the rotation angle is zero. Even rotating the tag, the recognition percentage does not go below the 20%. (b) A label (Tag 3) is compared with the image of a different non-rotated label (Tag 4) (i), rotated 90° (ii) and 180° (iii). The graph shows the recognition percentage that goes below the 1% when the rotation angle is zero. Rotating the tag, the recognition percentage continue to decrease. II level: fluorescence analysis by illuminating the tags with a light source at 370nm. (c) Tag 3 shows three fluorescence peaks with almost the same intensity (d) Tag 4 shows a more intense S420 fluorescence peak than Fluorescein and Rhodamine B. e) SEM image of the tag, f) Speckle analysis (III level) shows the FHD on this PUF typology, the "unlike" and "like" distribution centered at 0.5 and 0.14 respectively, then comparing two different PUFs the intra-device FHD results unrelated and centered at 0.5

Finally, the read-out time are different for the three encryption levels. As regards the first encryption level, the authentication procedure is very fast (2 seconds), reliable and within everyone's reach. Specific instrumentation is needed for the second and the third level authentication that, if available, allows to perform the procedure in less than 5 minutes. Moreover, in order to increase the encryption level of the proposed labels, another test has been conducted using a third fluorescent dye, Fluorescein. So, three different polymer/water solution, in the same percentage by weight, doped with different dyes were prepared: the PVA/water solution doped with S420 were used to produce fibers, while those doped with Fluorescein and Rhodamine B were used to produce both fibers and droplets. As expected, the results produce tags that are even more complex for all three security levels presenting a new encryption given by the third fluorescent dye. The tag possesses new emitted colors when stimulated with an external light source (see Figure 4a. As a consequence, the map of points presents an higher number of scatterers improving the security level, and on the other hand, a new emission peak appears in the fluorescence spectrum when the tags are illuminated with a light source at 370nm, as reported in Figure 4b,c respectively. The fluorescent emission of Fluorescein is stimulated directly with the light source that sets at 370nm, because it was difficult to obtain energy transfer between the three dyes using the light source at 310nm. Even if two of the used dyes do not have the maximum of absorption in the UV range a small fluorescence can still be stimulated. This is useful in terms of the proposed application because it allows the use of a single light source as for example a UV lamp. The presence of a third dye produces as a consequence a further increment of the uniqueness of the microscopic morphology. In fact, comparing two labels made with the same technique by means of the CRPs analysis, the intra-devices distributions result unrelated, the mean value of the FHD is centered at 0.5 (see green histogram in 4f). Here again to further confirm the analyzed behaviour the PCC for the proposed tag has been adopted and reported in Figure S4(a,b). In the SI Figure S5, we reported the analysis to set the threshold resolution (in terms of # of pixel flip in the challenge) to distinguish responses generated by different challenges. This threshold results equal to a flip of 1 px. Another important parameter to take into account is the encoding capacity (EC) of the proposed PUF systems which give information of the number of codes that can be generated. EC is defined as $k^N$, where $k$ is the number of bit contained in the responses (k = 2 for binary bits of 0 and 1) and $N$ is the number of independent bits. To this aim, the number of independent bits, performed by means of the binomial fit on the "unlike" distributions, $N = \mu\cdot(1-\mu)/\sigma^2$ where $\mu$ and $\sigma$ are the mean value and the standard deviation of that distribution respectively[36, 42, 45–47] for each analyzed PUF typology (produced using 1,2 and 3 dyes) is reported in Figure 5. This reveals that passing from the tag realized using solely the electrospinning to the mixed techniques one, the $N$ value increases. In fact the EC value passes from $2^N \sim 2^{400}$ in the first tag, to $2^N \sim 2^{640}$ in the second, and finally to $2^N \sim 2^{700}$ unique digits into the third one. While maintaining the same physical size of the challenge but increasing the number of macro pixels that composes it (i.e. 32x32 instead of 16x16) the unique digits EC increase to $2^N \sim 2^{1140}$ for the PUF with one dye, $2^N \sim 2^{1547}$, and $2^N \sim 2^{1725}$ for the PUF made by adding two and three dyes respectively. It is worth noting that these values are retrieved using the same parameters of the Gabor hash filter. In fact, the maximum number of independent bits for the Gabor hashed images, considering a wavelength of wavelet of 6 pixels, results N=2700 hence the maximum number of unique bits observed passes from 14% for tags with 1 dye illuminated with a 16×16 pxs challenge set to 60% for tags that involve 3 dyes and are illuminated with a challenge set composed by 32×32 pxs. The increasing $N$ value means that the dye inclusions perturb the tag morphology of each label. However, for each PUF typology, the encoding capacity remains almost constant within different tags representing a great advantage in cryptography application (electronic signature, communications, just to name a few) where usually 512 independent bit keys are used.

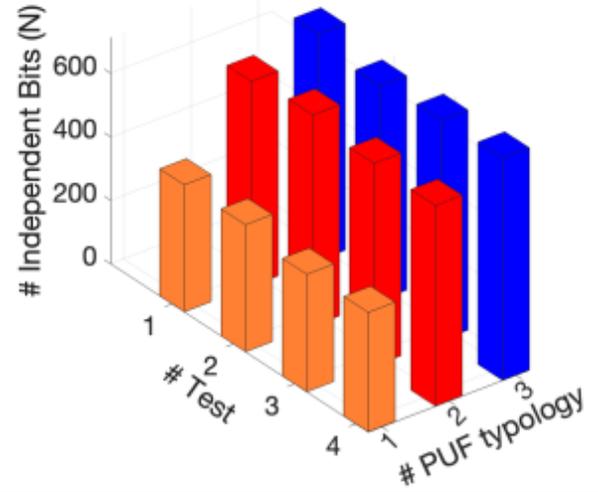

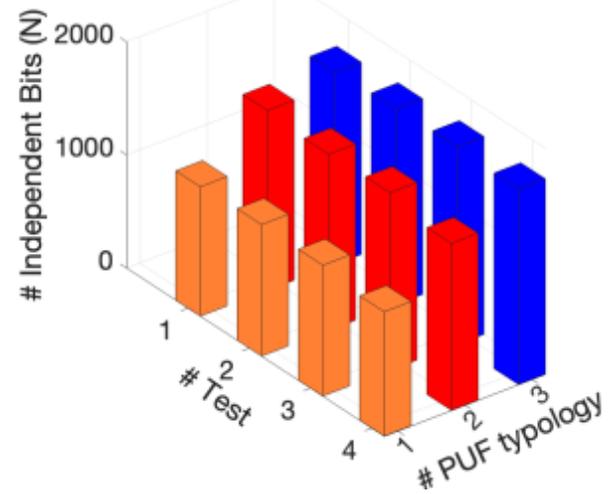

FIG. 5. Number of independent bits ($N$) for 4 tags of each typology (where 1,2 and 3 are referred to the number of involved dyes) a) using a challenge set input of 16×16 pixels and b) of 32×32.

**Stability tests**

Several tests were carried out to demonstrate the stability of each encryption level upon exposure to a UV light source for 1 hour and then to direct sunlight for 1 hour. The analysis was carried out on a label extracted from the batch fabricated using three fluorescent dyes, shown in figure 4. We will refer to this as Tag 5. Figure 6a shows the recognition percentage obtained for the authentication process of the map of scatterers carried out after UV light and sun exposure compared with the one before any exposure. There is a minor decrease in recognition percentage of points at 0 degrees, but the value is still above 60%. Similar values are obtained by rotating the label. Therefore, the first encryption level is guaranteed also after UV and sun exposure.
The fluorescence of the label was measured before UV exposure and then after 1 hour of exposure to 5mW UV light. Figure 6b shows that S420 fluorescence intensity is stable while Fluoresceine and Rhodamine B intensities decrease of about 10% with respect to the not irradiated sample. Nevertheless, the ratio between the Fluoresceine and Rhodamine B peaks remains constant, still lower with respect to the S420 peak. Then, it can be inferred that the fluorescence fingerprint of the label is preserved. The same experiment was carried out exposing the label under direct sunlight for 1 hour. The recorded intensity variation is similar to what observed for the UV exposure. The obtained results demonstrate that the qualitative evaluation of the fluorescence of the label, regarded as a second encryption level, is stable after a long UV and sun exposure. Finally, it is reported the aging test performed using UV light lamp during the CRPs acquisitions. Initially, data acquisition was carried out for one hour without UV light on the label collecting the produces responses, then we analyzed them to test the stability reporting their "like" distributions. After that, we illuminated the label with a UV light with a power of 20 mW, that corresponds to a solar irradiance ($\lambda$) of 0.17 $W/(m^2nm)$, for 10 minutes, acquired the CRPs, and illuminated again with UV light, in total for one hour. The same procedure was repeated with a UV light power of 50 mW, corresponding to a solar irradiance of 0.4 $W/(m^2nm)$. As reported in Figure 6 c, both the like distribution calculated with FHD or PCC are stable with the initial values meaning that the aging process does not affect the stability of the label itself. Also considering twice the standard deviation value (blur ribbon) the value still remains under a considered FHD threshold value of 0.2 or above a PCC threshold value of 0.8 demonstrating a great advantage for real applications.

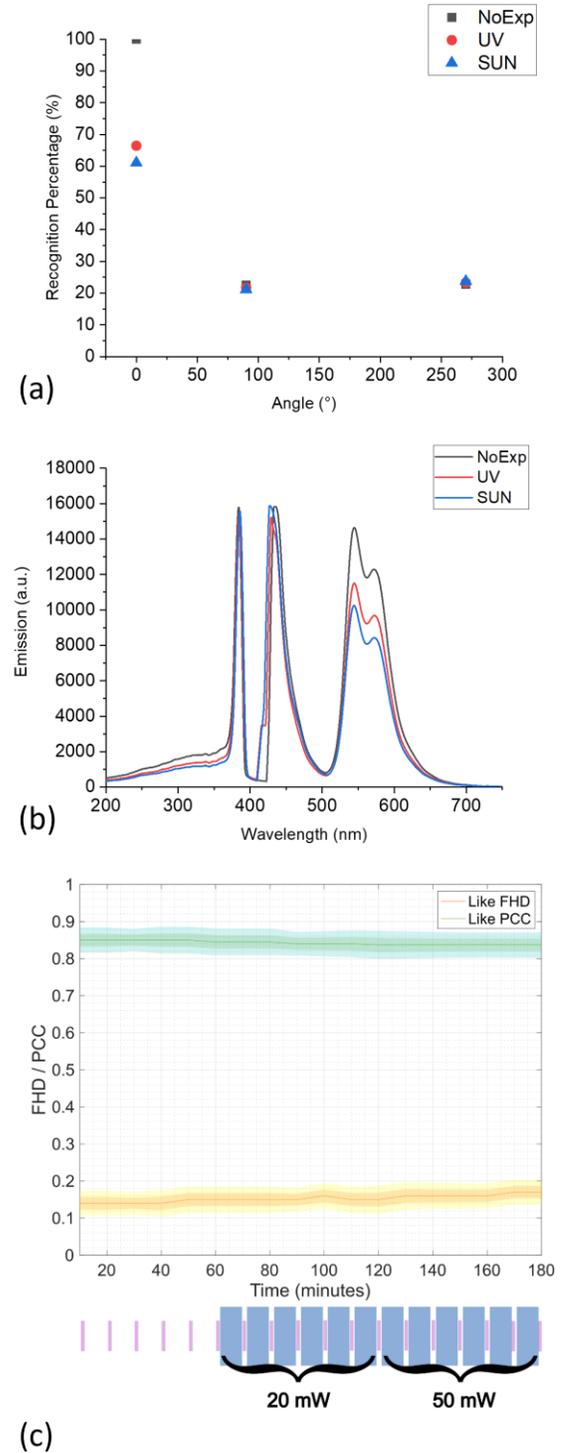

FIG. 6. Stability tests results after exposing the label to 1 hr of UV light and then 1 hr of sunlight (a) I level encryption recognition percentage (b) II level fluorescence intensity (c) III level CRP analysis.

## IV. CONCLUSIONS

This work presents a multi-level security label based on a unique distribution of scatterers, fluorescent emission, and speckle analysis, fabricated through electrospinning and electrospray techniques. The first level offers the possibility for the consumer to verify the authenticity of the product by taking a photograph of the label with a smartphone and verify it with a dedicated software. The fluorescence spectrum of the second security level can be analyzed both with a portable smartphone spectrophotometer or with a laboratory one. The third forensic level requires specific laboratory equipment and specialized personnel, and it is useful in lawsuits against forgers or for product control. The developed label presents characteristics such as non-clonability, ease of reading, low cost and eco-friendly feasible production. Overall, the proposed anti-counterfeiting labels –– thanks to their multi level security –– can be applied to many daily life products spanning from clothes to luxury goods, personal documents and for the certification of drugs and medical instruments.

## V. MATERIALS AND METHODS

**Samples fabrication:** Fabrication starts with the realization of a polyvinyl alcohol/water solution: 8% by weight of PVA (MW 85,000-124,000, 99%+ hydrolyzed, Sigma Aldrich) is added to water and the solution is stirred for 40 minutes at 90°C with a magnetic stirrer (Calctec). For each fluorescent dye a different PVA/water solution is prepared. Each dye doped PVA/water solution is prepared in the following percentages by weight: 99.7%*PVA/water*+0.3%*Dye*. The fluorescent dyes used are S420, Rhodamine B and Fluorescein (all from Exciton). Polymer nanofibers are realized using an homebuilt electrospinning setup composed of a high-voltage power supply, a syringe pump, a spinneret and a conductive collector plane. The nanofibers are collected on a collection plane. A voltage of 19–21 kV using a current generator is applied between the needle and the collection plane, that are placed 10–15cm one from the other. The solution is injected through a capillary with a flow rate of 1*ml/h*. Polymer microdroplets are produced using the same experimental setup but applying a voltage of 6 – 10 kV and increasing the flow rate to 2*ml/h*. The obtained label must have maximum thickness of 150 $\mu$m, that is a trade-off of mechanical stability and the optical thickness of the sample in order to still allow collecting CRPs with the CCD/ CMOS exposure time (20 ms) and an high frame rate (40 Hz).

**Characterization:**

**Map of Point recognition:** the analysis was carried out through a Java software application based on opensource image recognition tools [49]. A computer vision library implementing the Scale Invariant Transform (SIFT) algorithm [50] was exploited for pattern recognition in images acquired using mobile devices. SIFT algorithm allows us to elaborate 2D images without taking into account the scale zooming, rotation, and brightness changing. The detection of the distinctive features of an image consists of identifying scales and locations that can be assigned numerous times to the same object from different prospective. The analyzed images of the label possess a resolution of 96 dpi (954x1008).

**Fluorescence emission:** the measurements are performed using a spectrophotometer Avaspec-2048 (Avantes, U.S.A) using flexible optics fibers with a diameter of 500μm. The acquisition of the fluorescence spectra is made using a UV light source that emits at wavelength of 310nm and 370nm (Xenon lamp equipped with Omni$\lambda$ Monochromator) and a spectrometer equipped with an optical fiber placed at 10cm from the tag and at 45°. The UV lamp is able to stimulate at the same time all the fluorescent dyes used.

**CRPs experimental setup:** to obtain speckle patterns a red laser beam with wavelength $\lambda$=633nm (power 5 mW) propagates through a series of lenses, polarizers, and irises. The beam, after beam spot magnification by a beam-expander, impinges on a digital micro-mirror device (DMD) used for the challenge ($C_i$) generation by means of an intensity modulation. These challenges are generated using the "randi" Matlab function. It allows generating challenges that are unique between themselves. This function returns a pseudo-random scalar integer, for our scope between 0 and 1, that composes each challenge matrix, thanks to the use of a different seed in each run we obtain different challenges. The beam wave acquires individual spatial features for each individual $C_i$, that directly interrogate the scattering of the polymeric label illuminating an area of 1$cm^2$. The light interference produces a transmission optical pattern in the far field named speckle pattern. This constitutes the physical unclonable function (PUF) response $R_i$, that is collected in cross polarized configuration in order to remove any non-scattered light. This pattern is collected by a CCD camera. Here, we used a 270×360 px camera with 40 FPS for this task. All the details of the experimental setup are reported in [47, 48].

**Post processing:** Once the CRPs are collected they are filtered using a Gabor Hash filter. The Matlab function used to this end is:

[mag,phase] = imgaborfilt(A,wavelet, angle)

it computes the magnitude and phase response of a Gabor filter for the input gray scale image A. Wavelet describes the wavelength in pixels/cycle of the sinusoidal carrier, while angle is the orientation of the filter in degrees. To this end the following values are set: wavelet=6 pixels, angle = 45°, setting these values allows extracting correctly the features contained into each speckle image [51, 52]. In order to correctly evaluate these parameters we used the "xcorr2" Matlab function that allows self-correlating the collected pictures and evaluates the speckle grain size (filter wavelength) and orientation [53].

## VI. SUPPORTING INFORMATION

Supporting Information is available from the journal or from the authors.

## VII. AUTHOR CONTRIBUTIONS

M.D.L.B. conceived the idea and coordinated the overall research effort. M.D.L.B., G.E.L and A.Fe. carried out the experiments. G.E.L, S.N. and F.R. designed and implemented the experimental setup for the challenge response pairs scheme, they collected, studied and analysed the speckle data. A.Fo. and G.P. developed the app for mobile devices. G.D. carried out the analysis with SEM. M.P.D.S., G.G., D.S.W., R.C. and R.C.B. discussed the results. The manuscript was written by M.D.L.B., G.E.L. and A.Fe. with the input from all authors. The project was supervised by R.C.B.

## VIII. ACKNOWLEDGEMENTS

M.D.L.B., A.F, M.P.D.S., R.C. and R.C.B. thank project : "DEMETRA – Sviluppo di tecnologie di materiali e di tracciabilità per la sicurezza e la qualità dei cibi" PON ARS01 00401.
The LENS research group thanks the FASPEC (FiberBased Planar Antennas for Biosensing and Diagnostics), the project "Complex Photonic Systems (DFM.AD005.317). The publication was made with the contribution of the researcher G.E. Lio with a research contract co-funded by the European Union—PON Research and Innovation 2014-2020 in accordance with Article 24, paragraph 3a, of Law No. 240 of December 30, 2010, as amended, and Ministerial Decree No. 1062 of August, 2021.

**SUPPLEMENTARY INFORMATION**

To demonstrate that the only PVA/water solution doped with Rhodamine B it is not suitable for the realization of the proposed label, an example is reported in Figure S1. The solution is deposited on an aluminum foil using both electrospinning and electrospraying technique. Illuminating the obtained sample (Figure S1a) with a UV source of light(310nm) it is evident that the fluorescence is less bright than the PUF typology prepared using PVA/water solution doped with S420 to produce the base layer of fibers (Figure S1b,c).

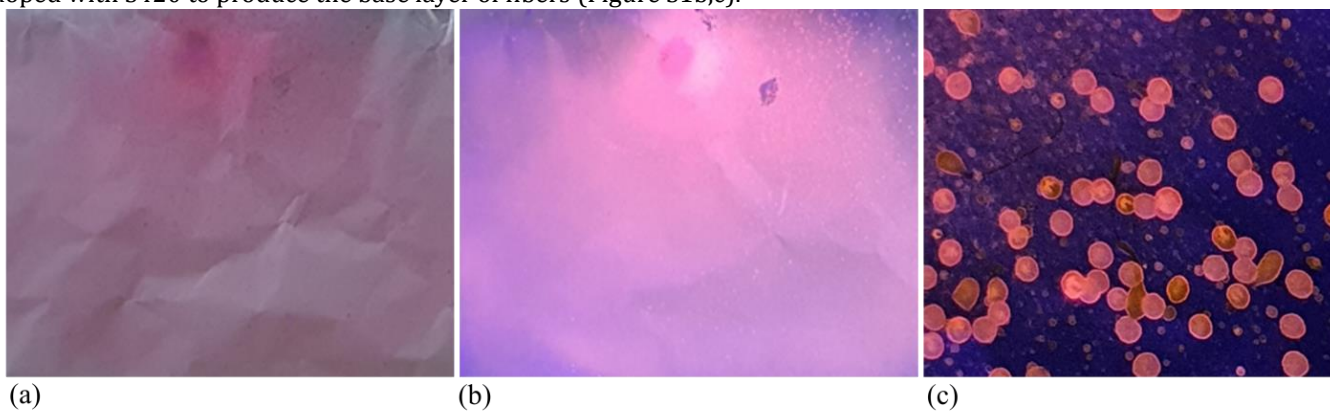

(a)          (b)          (c)

FIG. S1. (a) A PVA/water solution doped with Rhodamine B is used to produce polymeric fibers and droplets. (b) Illuminating the sample with a UV source of light emitting at (310nm), it does not show intense fluorescence emission. (c) A small area with large amount of droplets is reported, to underline the role of the background.

The recognition percentage is calculated from matrices created by the algorithm used for mapping the scatterers point. The matrix is reported in Figure S2a. For the batch based on two fluorescent dyes, five tags were analyzed. Each tag is compared with itself at different rotation angle and each picture rotated is compared with the other acquisition angles, as shown in Figure S2b. The same analysis were done for the PUF typology based on three fluorescent dyes, as shown in Figure S2a,b.

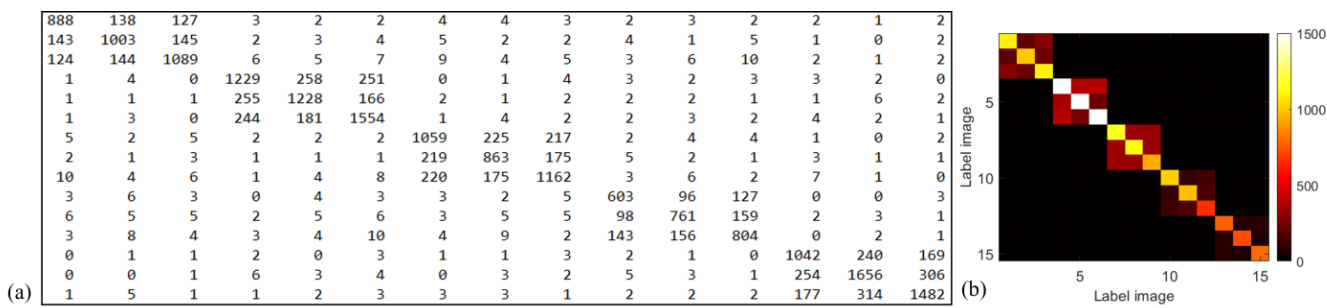

FIG. S2. (a) Matrix retrieved by using the developed algorithm for map of point recognition (b) relative correlation map.

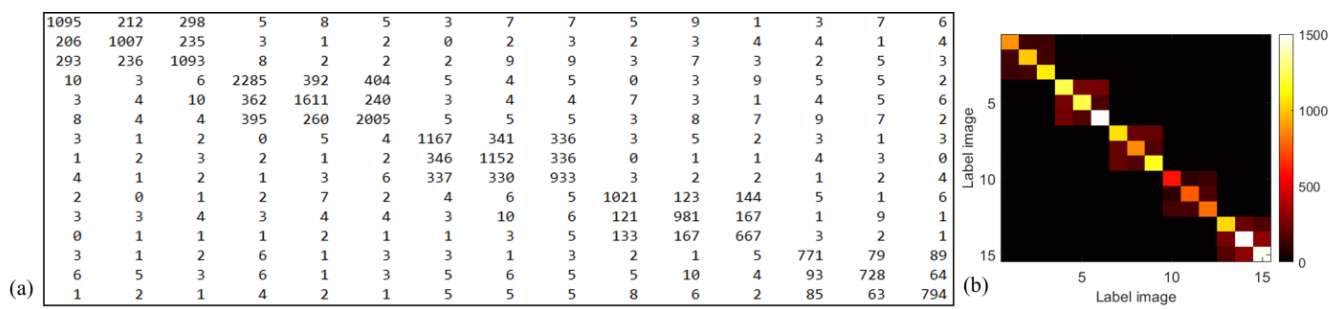

FIG. S3. (a) Matrix retrieved by using the developed algorithm for map of point recognition (b) relative correlation map.

The Figure S4 reports analysis about CRPs using the Pearson correlation coefficient (PCC) instead of FHD and Gabor hash filtering. The PCC analysis, that does not need any image process or hashing filtering of the responses, shows that for different responses generated with different challenges the "unlike" distributions goes close to 0, while for the "like" the PCC mean value goes close to 1, as reported in panel a, on the other hand, for two different tags the responses generated by the same challenges are uncorrelated ($<P> = \sim 0$). A further analysis has been conducted to evaluate the threshold resolution (# of pixel flip in the challenge) to distinguish two responses generated by two different challenges is 1 px. In particular it shows that the FHD distribution of the "like" responses and the ones interrogated by challenges that differ by only 1 px are well separated, see orange and green histograms in Figure S5a respectively. The separation between the two histograms increases accordingly to the number of involved dyes. These analysis has been also performed by using Pearson correlation coefficient without using any Gabor hash filter and response binarization as shown in Figure S5b.

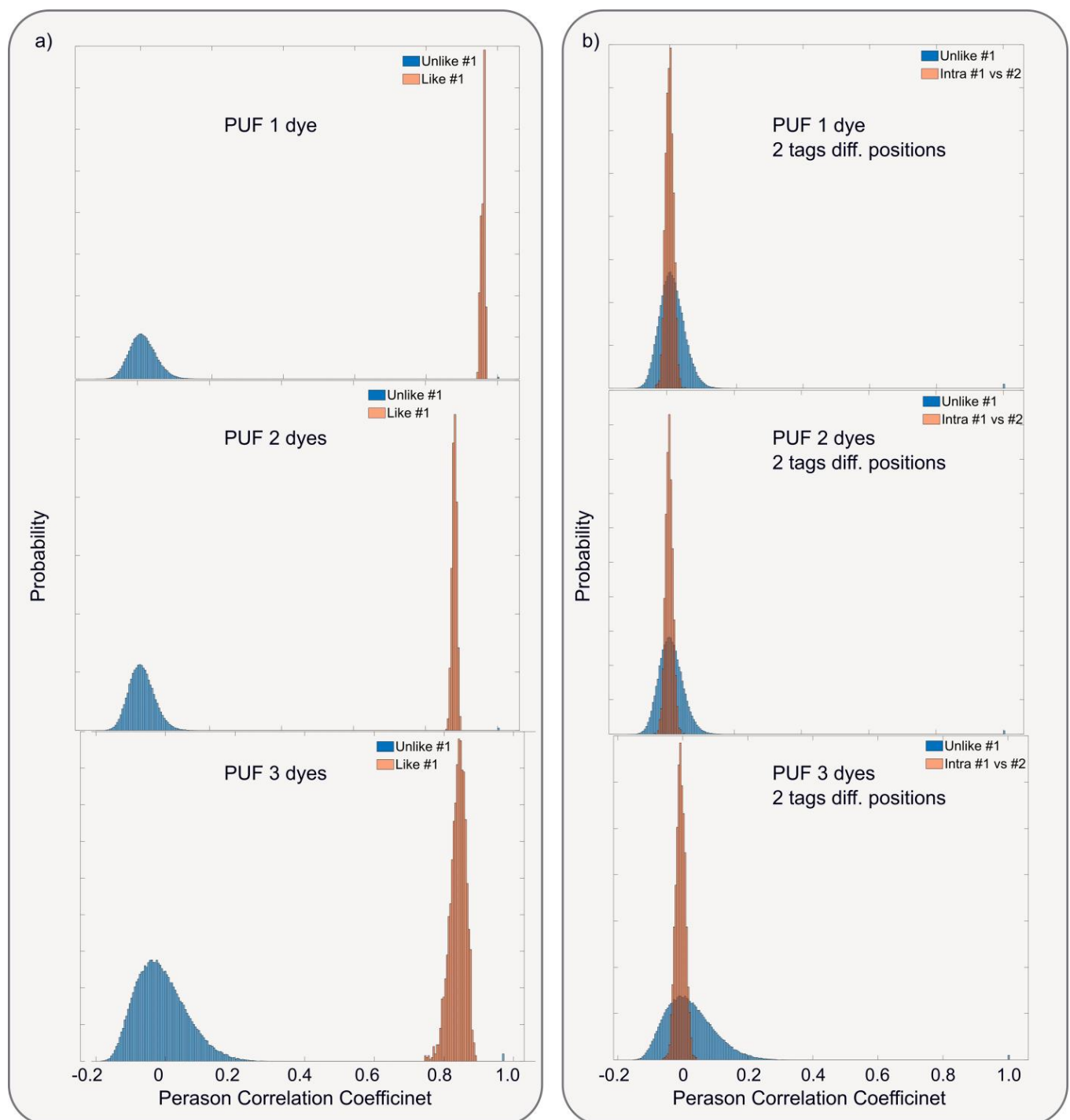

FIG. S4. a) Pearson Correlation Coefficient evaluated for responses generated with different challenges "unlike" and the "like" comparison. b) PCC for two different tags illuminated with the same challenge set. All these analysis have been performed on all PUF types.

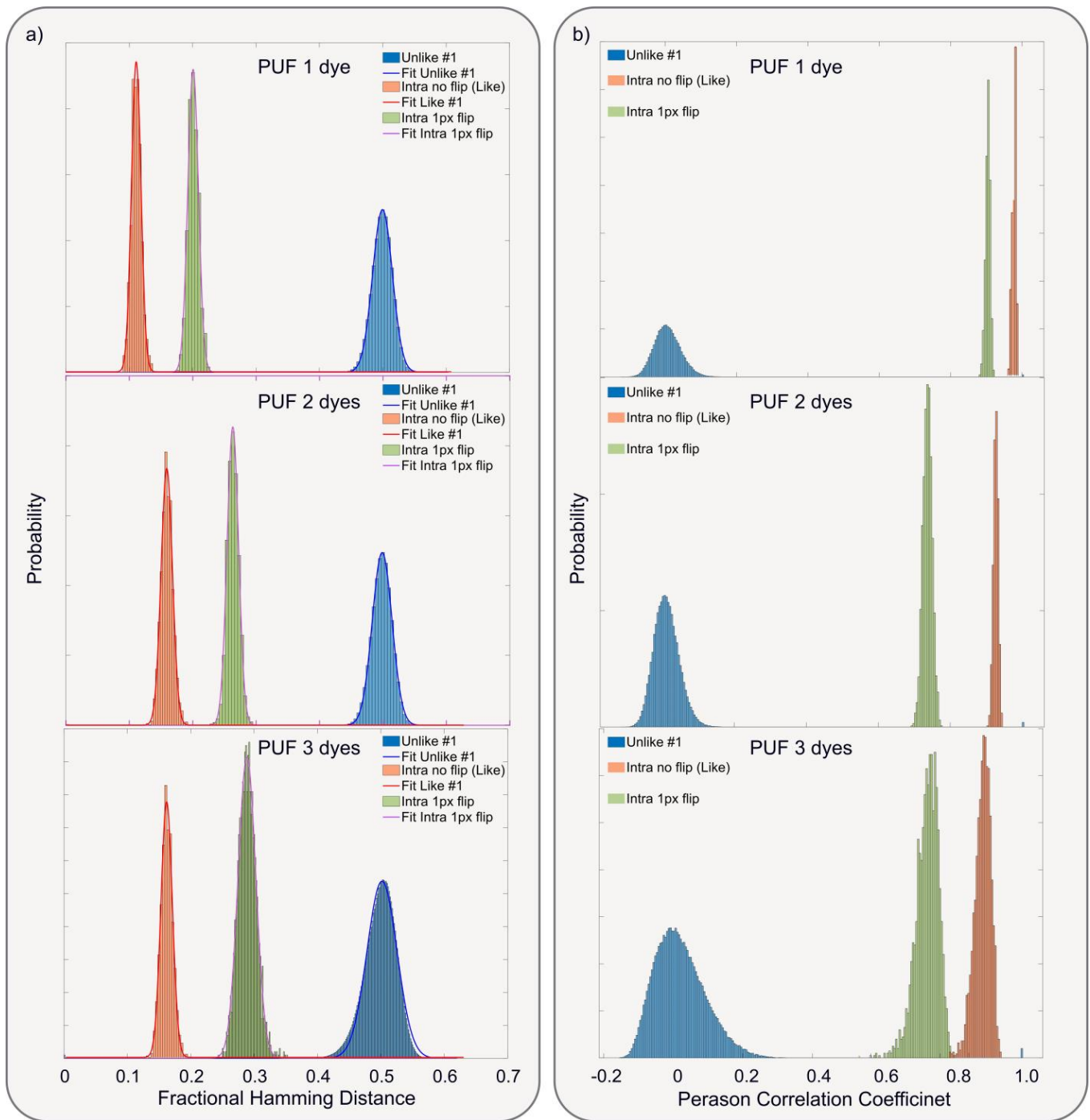

FIG. S5. (a)Fractional Hamming distance evaluated for all the PUF types considering the "unlike" distribution (blue histogram), the "like" ones, shown in orange, and the "intra 1px flip" shown in green. b) Pearson correlation coefficient evaluation for the same cases.